# Explainable Artificial Intelligence Reveals Novel Insight into Tumor Microenvironment Conditions Linked with Better Prognosis in Patients with Breast Cancer

Debaditya Chakraborty[1*], Cristina Ivan[2,3], Paola Amero[2], Maliha Khan[4], Cristian Rodriguez-Aguayo[2,3], Hakan Başağaoğlu[5], and Gabriel Lopez-Berestein[2,3]

[1]Department of Construction Science, The University of Texas at San Antonio, San Antonio, TX, USA. [2]Department of Experimental Therapeutics, The University of Texas MD Anderson Cancer Center, Houston, TX, USA. [3]Center for RNA Interference and Non-Coding RNA, The University of Texas MD Anderson Cancer Center, Houston, TX, USA. [4]Department of Lymphoma and Myeloma, The University of Texas MD Anderson Cancer Center, TX, Houston, USA. [5]Evolution Online LLC, San Antonio, TX, USA.

**\*Corresponding author:** Debaditya Chakraborty, Ph.D., Department of Construction Science, The University of Texas at San Antonio, San Antonio, TX, USA.
Email: debaditya.chakraborty@utsa.edu.

## Highlights

- New Explainable Artificial Intelligence (XAI) models predict survival in breast cancer patients.
- XAI relates the synergistic relationship between tumor microenvironment cells and patient survival.
- Novel insight into ideal B-cell and CD4+ T-cell fractions associated with better prognosis.
- B cells and CD4+ T cells play larger roles than CD8+ T cells in improving 5-year survival.

## Abstract

We investigated the data-driven relationship between features in the tumor microenvironment (TME) and the overall and 5-year survival in triple-negative breast cancer (TNBC) and non-TNBC (NTNBC) patients by using Explainable Artificial Intelligence (XAI) models. We used clinical information from patients with invasive breast carcinoma from The Cancer Genome Atlas and from two studies from the cbioPortal, the PanCanAtlas project and the GDAC Firehose study. In this study, we used a normalized RNA sequencing data-driven cohort from 1,015 breast cancer patients, alive or deceased, from the UCSC Xena data set and performed integrated deconvolution with the EPIC method to estimate the percentage of seven different immune and stromal cells from RNA sequencing data. Novel insights derived from our XAI model showed that CD4+ T cells and B cells are more critical than other TME features for enhanced prognosis for both TNBC and NTNBC patients. Our XAI model revealed the critical inflection points (i.e., threshold fractions) of CD4+ T cells and B cells above or below which 5-year survival rates improve. Subsequently, we ascertained the conditional probabilities of ≥5-year survival in both TNBC and NTNBC patients under specific conditions inferred from the inflection points. In particular, the XAI models revealed that a B-cell fraction exceeding 0.018 in the TME could ensure 100% 5-year survival for NTNBC patients. The findings from this research could lead to more accurate clinical predictions and enhanced immunotherapies and to the design of innovative strategies to reprogram the TME of breast cancer patients.

# 1 Introduction

Breast cancer is the most common cancer and the leading cause of cancer death in women worldwide. The prognosis is dependent on the type of breast cancer and on the stage of disease at detection [1, 2]. Breast cancer can be divided into several subtypes, based primarily on the expression of estrogen receptor (ER), progesterone receptor (PR), and human epidermal growth factor receptor 2 (HER2). Triple-negative breast cancer (TNBC) is a heterogeneous category of breast cancer, characterized by negative ER, PR, and HER2. TNBC is highly metastatic and aggressive, with poor prognosis, poor patient survival, and limited therapeutic options (3). Neoadjuvant chemotherapy, which is a combination of taxanes and anthracyclines, is the standard therapy for many TNBC patients. Whereas neoadjuvant chemotherapy is effective in some TNBC patients, about 50% develop resistance, leading to poor overall survival [3, 4]; hence, there is an urgent need for improved alternatives such as immunotherapies.

Cancer development, progression, and treatment resistance are known to be influenced by genetic and epigenetic alterations as well as by crosstalk between tumor cells and the tumor microenvironment (TME) [5]. The TME involves a complex network of soluble factors, tumor cells, and stromal cells that play a crucial role in the initiation, development, and progression of breast cancer. The TME in breast cancer consists of multilineage immune cells (e.g., T and B lymphocytes, myeloid cells, and dendritic cells), cancer-associated fibroblasts, and tumor endothelial cells [6]. All of these cell subtypes live in an ocean of hormones, growth factors, and cytokines in the breast TME [7]. Adding to this complexity is a myriad of pathways that dictate the fate of the tumor, metastases, and patients' lives. The balance between tumor-infiltrating immune effector cells in the TME (such as CD4+ T cells and CD8+ T cells [or cytotoxic T lymphocytes, CTLs]) regulates the immune response cytotoxic effects on tumor cells. In contrast, tumor-infiltrating myeloid cells, such as tumor-associated macrophages, promote the expansion and dissemination of cancer cells depending on their functional state [8]. TME crosstalk potentially promotes cancer progression and TME plasticity. Adaptations to TME factors may be responsible for metastasis and immune evasion.

Multiple targeted therapies have been used for TNBC but to no avail. Aberrant signaling by VEGFR2 and cMET, as well as modifications of the immune cell population in response to an immune-suppressive phenotype, has been shown to lead to the failure of immunotherapy of TNBC [9]. In TNBC, multiple genomic instabilities and mutations have been associated with immune responses [10]. A comparative study between TNBC and non-TNBC (NTNBC) showed that TNBC is characterized by higher expression levels of functional gene sets associated with 15 types of immune cells [11, 12]. Unfortunately, innovative strategies to reprogram the TME of breast cancer patients is challenging, in part because of conflicting findings in the literature. For example, tumor-infiltrating B lymphocytes (B cells) have been associated with positive, negative, or no significance in breast cancer prediction and prognosis [13]. Other studies have reported that targeting regulatory B (Breg) cell activity may be used to enhance immunotherapeutic outcomes [14].

Augmentation of CTL-induced antitumor immune reactions has been considered to be an attractive therapeutic modality for lethal solid tumors due to the tumor-killing ability of CD8+ CTL [15]. In the adaptive immune system, T-helper cells (CD4+ cells) play a critical role in releasing cytokines and primarily help CD8+ CTL and antibody responses to mediate antitumor immunity. However, the interplay between polyfunctional CD4+ T cells and other immune cell lineages within the context of tumor immunity is not well understood [16]. In addition to immune cells, tumor-associated factors in the TME have also been targeted in cancer therapies. The recognition that tumor-associated endothelial cells and cancer-associated fibroblasts are important mediators of immune suppression have led to the development of cell-specific targeting drugs in an attempt to enhance the immune response [6]. Tumor-associated macrophages are able to suppress the functions of CD8+ T and NK cells and promote tumor cell growth in the TME [17].

We provided a new perspective through an entirely data-driven artificial intelligence (AI) approach toward clinical cancer research. AI could lead to a paradigm shift in cancer treatment, thereby resulting in major improvements in patient survival due to enhanced predictions [18]. In this article, we develop Explainable Artificial Intelligence (XAI) models to establish and investigate the data-driven relationship between TME features, including CD4+ and CD8+ T cells, B cells, endothelial cells, fibroblasts, and macrophages, and the overall and 5-year survival rates of both TNBC and NTNBC patients. The XAI models also assign relative influence of immune cells (T cells, B cells) and tumor-associated cells—including fibroblasts, endothelial cells, and macrophages—in the TME on the overall and 5-year survival rates of TNBC and NTNBC patients. In addition, using XAI models and conditional probabilities, we identified and analyzed the inflection points of the critical microenvironment features above or below which the ≥5-year survival rates could potentially improve in both TNBC and NTNBC patients. The resulting new perspective on favorable or deleterious microenvironmental conditions could lead to improved prognoses through well-informed clinical management and therapeutics, including the design of innovative strategies to reprogram the TME of breast cancer patients.

## 2 Materials and Methods

### 2.1 Patients and databases

We downloaded clinical information from TCGA about patients with invasive breast carcinoma (BRCA) from two studies from the cbioPortal (http://www.cbioportal.org/), the PanCanAtlas project and the GDAC Firehose project [19, 20]. For this cohort, we downloaded from UCSC Xena (http://xena.ucsc.edu/) normalized RNA sequencing data in transcripts per million (TPM) units [21]. Furthermore, we used the R package immunedeconv (https://icbi-lab.github.io/immunedeconv/) to perform integrated deconvolution based on the EPIC method to estimate the percentages of seven different immune and stromal cells (B cells, cancer-associated fibroblasts, CD4+ T cells, CD8+ T cells, endothelial cells, macrophages, NK cells) from RNA sequencing data [22, 23].

### 2.2 Explainable Artificial Intelligence (XAI) model

Over the past decade, there has been a major increase in the number of large and complex omics datasets [24, 25], especially through large consortium projects such as TCGA, which has sampled multiomics measurements from more than 30,000 patients and dozens of cancer types [26]. These rich omics data provide unprecedented opportunities to systematically characterize the underlying biological mechanisms involved in the evolution of cancer and to understand how the TME (stromal cells, immune cells, and other types of cells) contributes to this evolution [25, 27].

Novel techniques in AI can bring together diverse data types to expand novel insights gained from the multiomics datasets. It is well acknowledged that enrichment of high-quality data coupled with machine learning, a subset of AI, can help identify areas of changing patients' unhealthy behaviors [28], risk prediction or recurrence prediction for chronic diseases after a curative treatment [29], therapeutic need, enhance clinical trial interpretation, and even identify novel targets [30]. However, a major criticism of incorporating AI, particularly deep learning, into medical fields is the idea that AI is essentially an opaque "black-box" that is mechanistically uninterpretable [31, 32, 33]. The assumed lack of interpretability of AI models has been a debated topic within the field, with models cited that have achieved high accuracy due to factors that are not useful in prospective predictions [31, 34, 35]. The crux of the problem is that linear models, although interpretable, produce less accurate models when the datasets are complex and inherently nonlinear. In such cases, tree-based ensemble models, which are interpretable models [29], can be used in lieu of deep learning models that allow scientists and clinicians to understand the underlying reasoning behind the decisions and predictions. Recently, Gu et al. [36] used a tree-based ensemble model, called extreme gradient boosting (XGBoost) [37], to predict the risk of breast cancer relapse from clinical data (e.g., age, tumor size, treatment) and then use case base reasoning - which solves new problems by constructing a historical case base and using the results of similar historical cases - to explain the reason for the prediction. In addition, the authors used a game theory based SHapley Additive explanation model called SHAP [38, 39] for global explainability of the results to identify the order of importance of the clinical features considered. In this paper, we developed data-driven XAI models using XGBoost and SHAP to enhance the explainability of the breast cancer survivability models based on TME conditions (including both immune cells and tumor-associated cells) and understand the underlying reasoning and expand our knowledge without compromising predictive accuracy. Moreover, in addition to the global SHAP analysis to determine the order of importance of the TME cells on the patients' survivability rates, we performed local SHAP analysis to identify the inflection points for the TME cells at above (or below) the survivability rates may increase. We demonstrated that the local SHAP analysis expanded the potential use of the interpretable AI model to investigate potential immunotherapies to increase patients' survivability rates with enhanced explainability and transparent reasonings.

XGBoost is a variant of a tree-based boosting algorithm. Conceptually, XGBoost learns the functional relationship $f$ between the features $x$ and target $y$ through an iterative process in which the individual trees are sequentially trained on the residuals from the previous tree. Mathematically, the predictions from the trees can be expressed as

$$\hat{y} = \phi(x) = \frac{1}{n} \sum_{k=1}^{n} f_k(x),$$  Eq. 1

where $\hat{y}$ is the predicted outcome (overall survival and 5-year survival) in breast cancer patients, $1 \leq k \leq n$, and $f_1, f_2, \ldots, f_n$ are the functions learned by the $n$ number of trees.

The following regularized objective $\mathcal{L}(\phi)$ is minimized to learn the set of functions $f_k$ used in the model:

$$\mathcal{L}(\phi) = \sum_i l(\hat{y} - y) + \sum_k \Omega(f_k),$$  Eq. 2

where $\Omega(f_k) = \gamma T + \frac{1}{2} \lambda ||w||^2$.

In Eq. 2, $l$ is the differentiable convex loss function that measures the difference between $\hat{y}_i$ and $y_i$. $\Omega$ is an extra regularization term that penalizes the growing of more trees in the model to prevent complexity and thus, reduce overfitting. $\gamma$ is the complexity of each leaf, $T$ is the number of leaves in a tree, $\lambda$ is a penalty parameter, and $||w||$ is the vector of scores on the leaves. Note that if the regularization parameter $\Omega$ is set to zero, the objective falls back to the traditional gradient tree boosting.

SHAP, in contrast, was used to explain the AI models, that is, to investigate the relationship and contribution of each feature to the predicted AI-based outcome ($\hat{y}$). SHAP computes the Shapley values that signify the average marginal contribution of each feature value across all possible combinations of features. The features with large absolute Shapley values are deemed impactful. To evaluate the overall feature influence on the predicted outcome, SHAP averages the absolute Shapley values for every feature across the data, sorts them in decreasing order, and plots them. In our work, negative Shapley values associated with the feature instances indicate better chances of overall and ≥5-year survival.

## 3   Results

We developed two sequential models with the XAI pipeline (schematic in Fig. 1): (i) Model A, which predicts the probability of overall survival of patients with breast cancer based on percentages in the TME of B cells, CD4+ T cells, CD8+ T cells, endothelial cells, macrophages, and fibroblasts and on ER/PR/HER2, age at diagnosis, and cancer stage; and (ii) Model B, which predicts the probability of 5-year survival for both TNBC and NTNBC patients based on the output from Model A in addition to the input features used in Model A. The XAI models were developed through a multistep process shown in Figure 1, which includes (I) data preprocessing steps, (II) hyper-parameter optimization via a 3-fold cross validation to find the best subset of hyperparameters, (III) testing of the predictive accuracy of the final AI models after being trained with the best subset of hyperparameters, (IV) predicting the probability of the clinical outcomes, and (V) explanation of the predicted outcomes from Model B with a game theory–based XAI model to reveal the contributing factors and respective values of the TME constituents that lead to better outcomes (≥ 5 years survival) for breast cancer patients. We quantified the conditional probability of ≥5-year survival ($S_5$) given a certain TME condition ($C$) using:

$$P(S_5|C) = 100 \times \frac{P(S_5 \cap C)}{P(C)}, \quad \text{Eq. 3}$$

where $P(S_5 \cap C)$ is the probability that both events $S_5$ and $C$ occur simultaneously, and $P(C)$ is the probability of the condition ($C$) to occur.

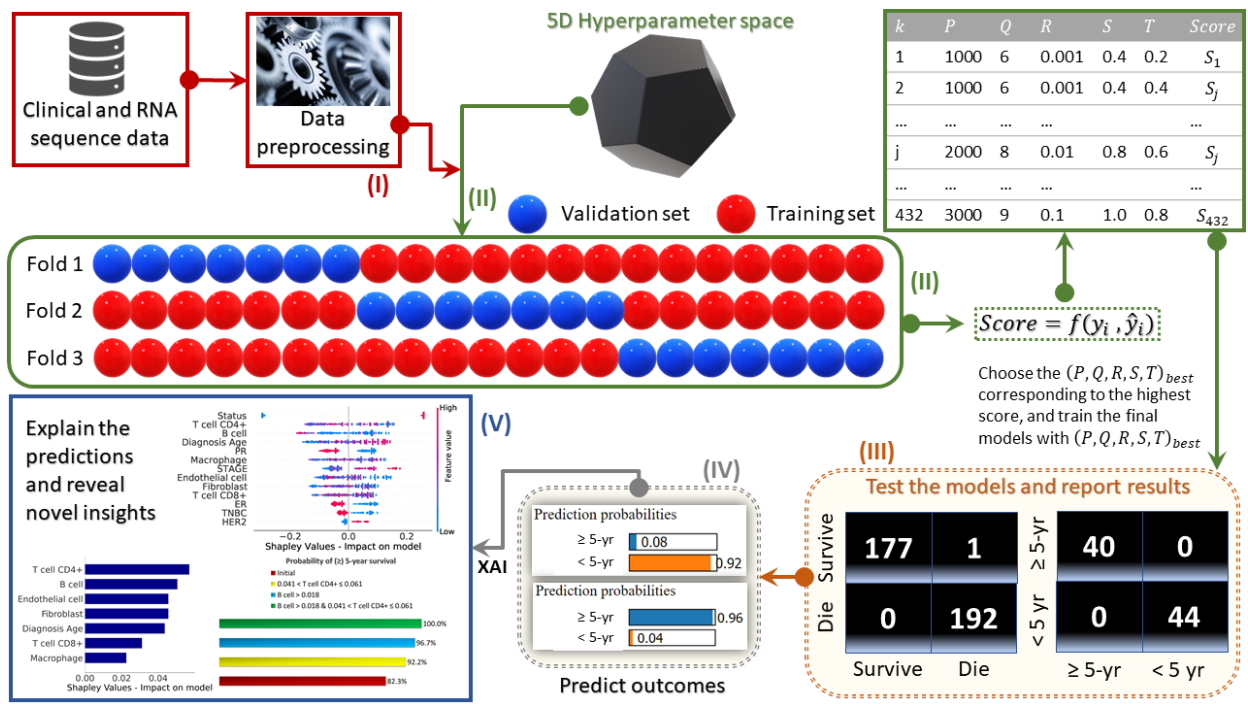

**Figure 1:** Schematic representation of the modeling pipeline: (I) data preprocessing steps, which include encoding categorical features as integer arrays, range-normalization of immune and tumor cell data for enhanced XAI interpretability, and data oversampling to convert from an imbalanced to a balanced dataset (i.e., the numbers of surviving and deceased patients were made nearly equal) to avoid bias (i.e., preventing the model from ignoring the minority class); (II) hyperparameter optimization via a 3-fold cross-validation to find the best subset of hyperparameters ($P$- number of estimators, $Q$- maximum depth of each estimators, $R$- learning rate, $S$- subsample ratio, $T$- column sample ratio for each estimator) that improves the models' roc-auc score $\{f(y_i, \hat{y}_i)\}$ signifying the area under the receiver operating characteristic curve during the (432×3) iterations over the hyperparameter space (i.e., to enhance the predictive accuracy of the XAI model); (III) testing the predictive accuracy of the final AI models after being trained with the best subset of hyperparameters; (IV) predicting the probability of the clinical outcomes (overall and 5-year survival rates); (V) explanation of the predicted outcomes with a game theory–based XAI model to enhance the interpretability and explainability of the model predictions, identification of critical inflection (turning) points above (or below) which the 5-year survival rates increase, and assessing the conditional probability of ≥5-year survival rates, given that the TME factors are within certain ranges determined by the inflection points.

With any data driven XAI model, it is imperative to ensure that the model produces accurate predictions on samples that were not used during model training and that the relevant statistical measures obtained during model training and testing are comparable to avoid overfitting or underfitting the data. We analyzed and reported the confusion matrices (Fig. 2) to gain a better understanding of the models' performance in predicting the overall (Fig. 2a and 2b) and 5-year (Fig. 2c and 2d) survival on both the training and testing data that constitute 75% and 25% of data, respectively, randomly sampled from the entire preprocessed dataset. These confusion matrices show that the off-the-shelf AI model built with an off-the-shelf XGBoost algorithm produces many false negatives, that is, patients who are "likely to die" were inaccurately predicted as "likely to survive," which is undesirable. The custom AI models, however, along with the randomly oversampled dataset, developed with the proposed pipeline produce reliable predictions on both the training and testing data (Fig. 2b and 2d). The accuracy, precision, recall, and F1 score of the proposed AI models on the testing data were 99.73%, 99.48%, 100%, and 99.74%, respectively, with regard to predicting overall survival (Fig. 2b) and were 100% (Fig. 2d) on the testing data with regard to predicting 5-year survival.

These results indicate that the custom AI models are capable of accurately predicting outcome for breast cancer patients based on the data associated with TME conditions in tumors and on the hormone-receptor statuses of the cancer. After evaluating the predictive ability of the models, we used XAI to interpret the custom model's predictions, investigate novel relationships between the TME features and the 5-year survival status, and identify the critical inflection points above or below which the 5-year survival rates improve. We explain the custom XAI models from both global (entire dataset) and local (individual data points) perspectives. The global explanations revealed that the CD4+ T cells and B cells are the most influential TME factors in determining the likelihood of overall survival (Fig. 3a) and 5-year survival (Fig. 3b) for breast cancer patients. XAI model A suggests that a very high or low CD4+ T-cell count, along with low

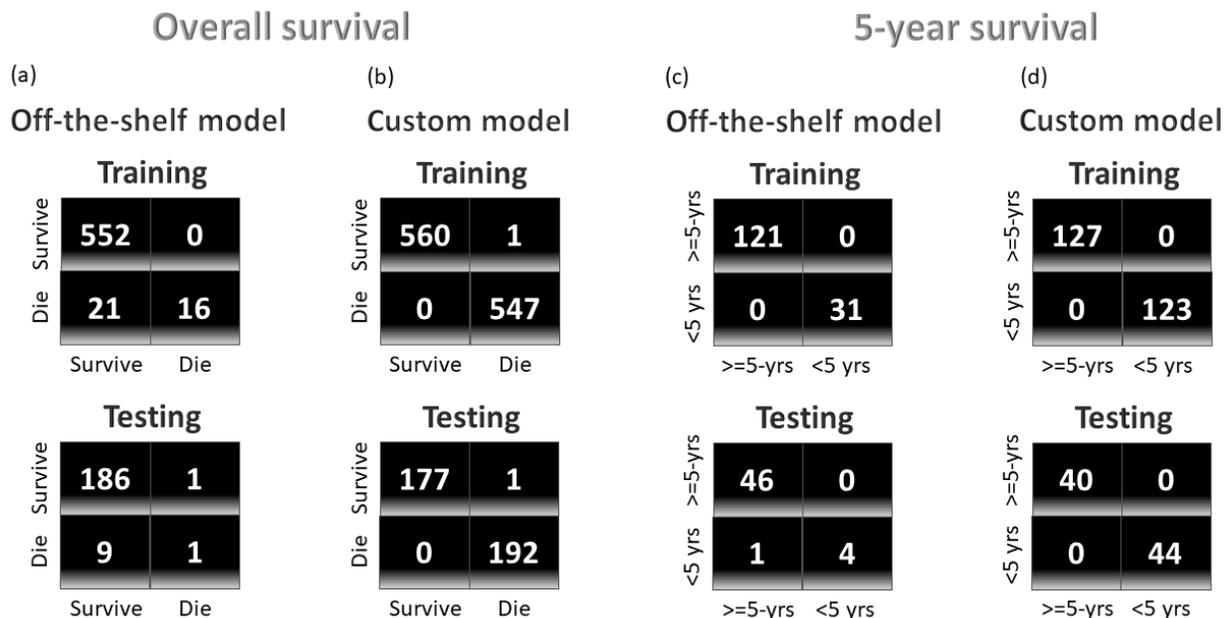

**Figure 2:** Model accuracy evaluation with confusion matrices on both training and testing data. Predictive accuracies of an off-the-shelf **(a)** and our custom machine **(b)** learning models in terms of their ability to predict the likelihood of overall survival of breast cancer patients. Predictive accuracies of an off-the-shelf **(c)** and our custom machine **(d)** learning models in terms of their ability to predict the probability of 5-year survival of breast cancer patients.

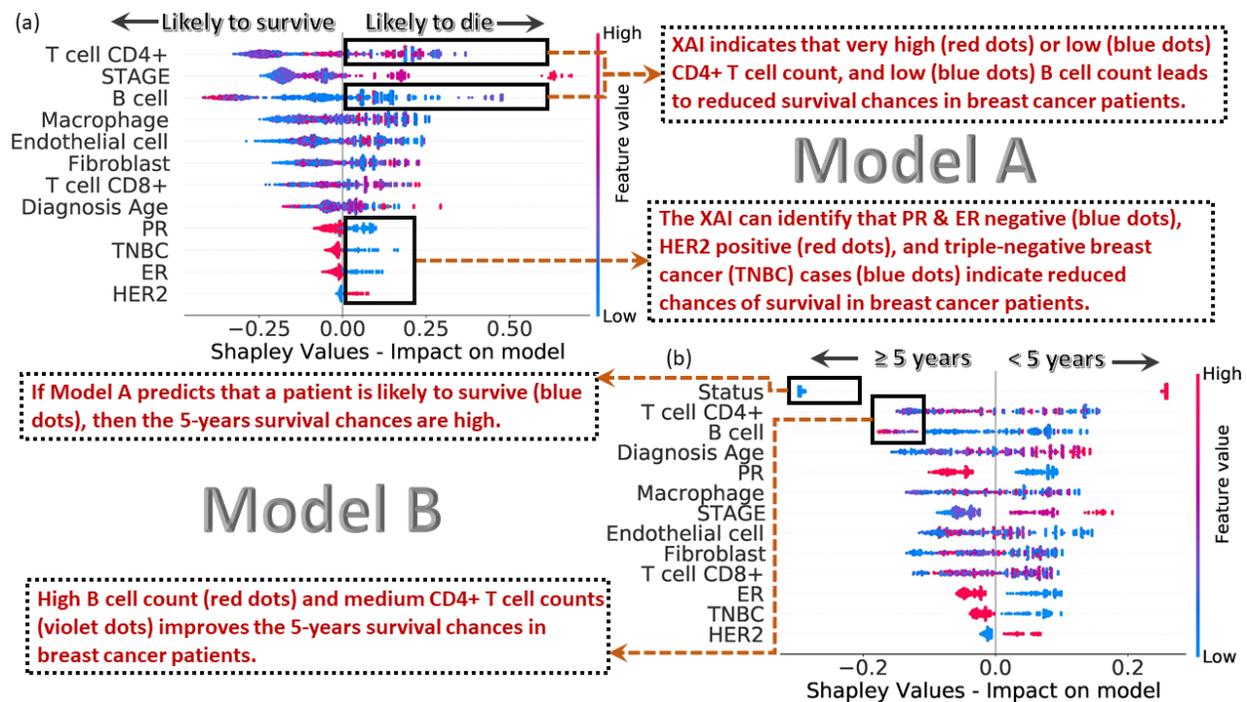

**Figure 3:** XAI results obtained from Models A **(a)** and B **(b)**, in which Model A predicts the likelihood of overall survival of breast cancer patients and Model B predicts the 5-year survival rates. The features on the y-axis are the independent variables used in the respective models; their relative positions were determined by their relative importance in making correct predictions. The XAI indicates that CD4+ T-cell and B-cell fractions are the most important features in the TME in terms of predicting outcome for breast cancer patients. Note: The blue dots represent lower feature values, and the red dots represent higher feature values. For example, blue dots corresponding to PR, ER, and HER2 represent PR-negative, ER-negative, and HER2-negative cases, respectively. Blue dots corresponding to TNBC represent triple-negative breast cancer cases (TNBC is encoded as 0, and NTNBC as 1). The status in (b) represents the output from Model A, where "likely to survive" is encoded as 0 (blue dots) and "likely to die" is encoded as 1 (red dots).

B-cell counts, leads to reduced survival rates for breast cancer patients. Model A also showed that PR- and ER-negative, HER2-positive, and TNBC cases were associated with reduced survival rates for breast cancer patients.

Interestingly, CD4+ T cells and B cells play larger roles than do CD8+ T cells in improving 5-year survival rates, as seen from the lower importance assigned by the XAI to the CD8+ T cells in terms of influencing outcome for breast cancer patients. In other words, stimulation and activation of CD4+ T-helper cells (that release cytokines to activate B cells and other immune cells in the TEM) and the activation of B cells (antibody-producing machines) appear to have more influential effects than do the CD+8 T cells with cytotoxic ability in breast cancer progression. Moreover, although the role of B cells was reported to be controversial in cancer immunotherapies [13], Figure 3a and 3b reveals that the presence and activation of higher numbers of B cells (presumably activated by the CD+4 T cells) in the TME could enhance patients' survivability rates and the efficacy of cancer immunotherapies. These findings are particularly important if the cancer is not at an advanced stage, at which a patient would be at higher risk of death according to Shapley analysis in Figure 3a. The analysis further indicates that activation and stimulation of T-helper cells (e.g., through chemical signaling induced by releases of cytokine) to alert, stimulate, and activate other immune cells in the TEM against tumor initiation or progression could be the most critical process to enhance the patient's chances of survival. These new findings and insights indicate an urgent need to rethink the current cancer immunotherapies that are largely focused on harnessing the antitumor CD8+ cytotoxic T-cell response [12, 16]. Because tumor-associated macrophages, endothelial cells, and fibroblasts were found to be less influential than CD4+ T cells and B cells, the augmentation of CD4+ T and B cell–induced antitumor immune reactions would be more effective than targeting cancer-associated factors in the TME in breast cancer therapies.

Since CD4+ T cells and B cells are the two most important TME features, according to the global explanations in Figure 3, we explain their effect on the model's predictions from a local perspective to identify the critical inflection points above (or below) which the 5-year survival rates improve (Fig. 4). We found that the CD4+ T-cell count within the range of 0.2 to 0.3 (Fig. 4a) (corresponding to 0.041 to 0.061 on the original scale [Fig. 4c]) is ideal for better outcomes (i.e., improved chances for ≥5-year survival) for breast cancer patients. Similarly, we ascertained that the ≥5-year survival rates were higher when the B-cell count was approximately >0.09 (Fig. 4b) on the normalized scale, which corresponds to 0.018 on the original scale of the data (Fig. 4c). We designed three different TME conditions based on these inflection points:

(i)     $C_1$: 0.041 < CD4+ T cells ≤ 0.061,
(ii)     $C_2$: B cells > 0.018, and
(iii)     $C_3$: B cells > 0.018 and 0.041 < CD4+ T cells ≤ 0.061,

which were coupled with Eq. 3 to quantify the conditional probability of ≥5-year survival ($S_5$) of breast cancer patients (Fig. 4d). We found that the initial probability of ≥5-year survival ($S_5$) based on the original dataset was 82.3%. In contrast, the probability of ≥5-year survival ($S_5$) given conditions $C_1$, $C_2$, and $C_3$, i.e., $P(S_5|C_1)$, $P(S_5|C_2)$, and $P(S_5|C_3)$ increased to 92.2%, 96.7%, and 100%, respectively. In other words, the augmentation of CD4+ T cells alone or B cells alone in antitumor immune reaction could boost the ≥5-year survival rate by 9.9% and 14.4%, respectively. When the augmentation of CD4+ T cells and B cells were analyzed together, the ≥5-year survival rate increased by 17.7%. The XAI-based revelation of the critical inflection points along with the statistical evaluation of certain TME conditions has high potential in deriving novel insights that could have clinical implications for accurate predictions and targeted clinical treatment of patients with breast cancer.

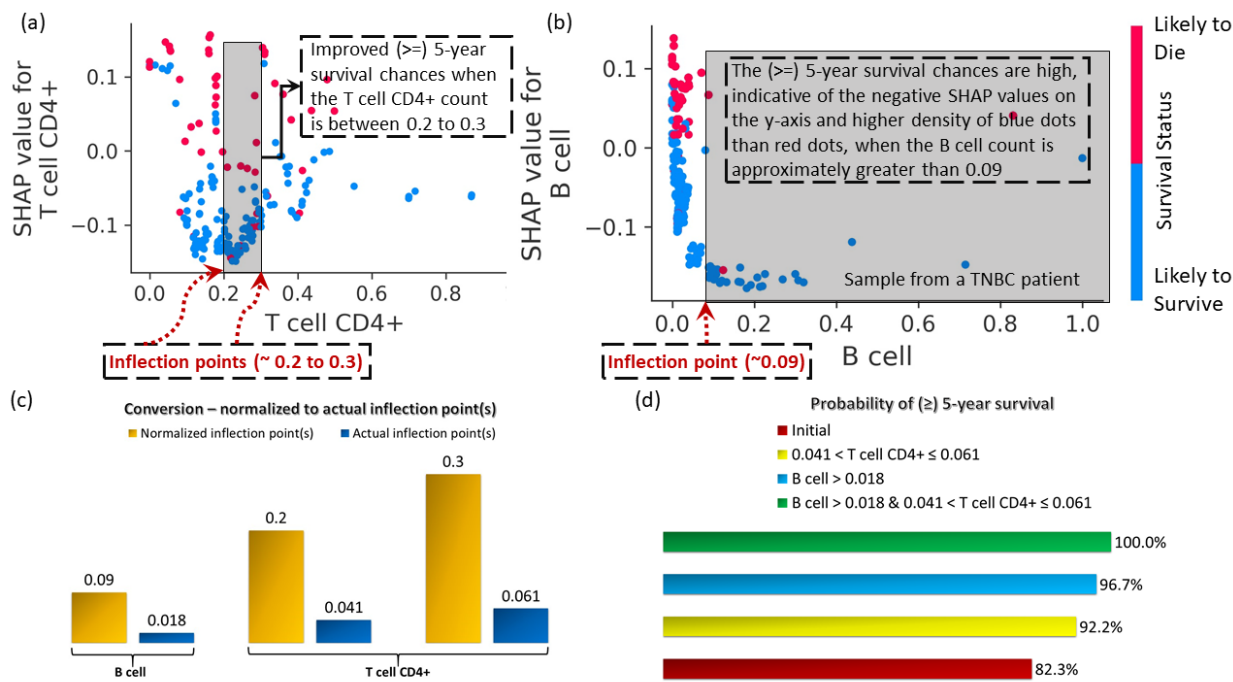

*Figure 4: XAI results based on data from breast cancer patients who survived ≥5 years and are still alive, are dead after surviving ≥5 years, or are dead after surviving <5 years. (a) and (b) reveal the interaction between CD4+ T cells and B cells with the ≥5-year survival rates, respectively. Lower SHAP values on the y-axis indicate higher chances of ≥5-year survival in breast cancer patients, and the optimal ranges of B cells and CD4+ T cells that could potentially improve patient outcome are shaded in gray, with inflection points marked. The conversion from the normalized inflection points to the actual inflection points based on the original scale of the data before conversion for XAI modeling is shown in (c). The conditional probabilities of ≥5-year survival of all breast cancer patients in various TME conditions from the clinical dataset of patients are shown in (d).*

We assessed the conditional probability of ≥5-year survival separately on patients with NTNBC and those with TNBC (Fig. 5a and 5b), which reaffirmed that conditions $C_1$, $C_2$, and $C_3$ improve prognosis in both cohorts. In fact, for NTNBC patients, having a B-cell fraction exceeding 0.018 could ensure 100% 5-year survival rates. We found that the most impactful factors for predicting the final outcome in TNBC patients who survived ≥5 years were CD4+ T cells and B cells (Fig. 5c); in TNBC patients who survived <5 years, cancer stage and age at diagnosis, as well as a lower impact of CD4+ T cells and B cells, influenced final outcome (Fig. 5d). Additional statistical analysis of the data in Figure 5d revealed that the median conditional probabilities for CD4+ T cells and B cells in TNBC patients who survived <5 years were 0.037 and 0.004, respectively, which were lower than the ideal range of these immune cells identified by our XAI model. Thus, the findings in this study revealed that CD4+ T cells and B cells play a much more significant role than do other TME factors in terms of survival, and the importance of our focus on maintaining the ideal CD4+ T cell and B cell fractions may lead to better prognoses, even for patients with TNBC.

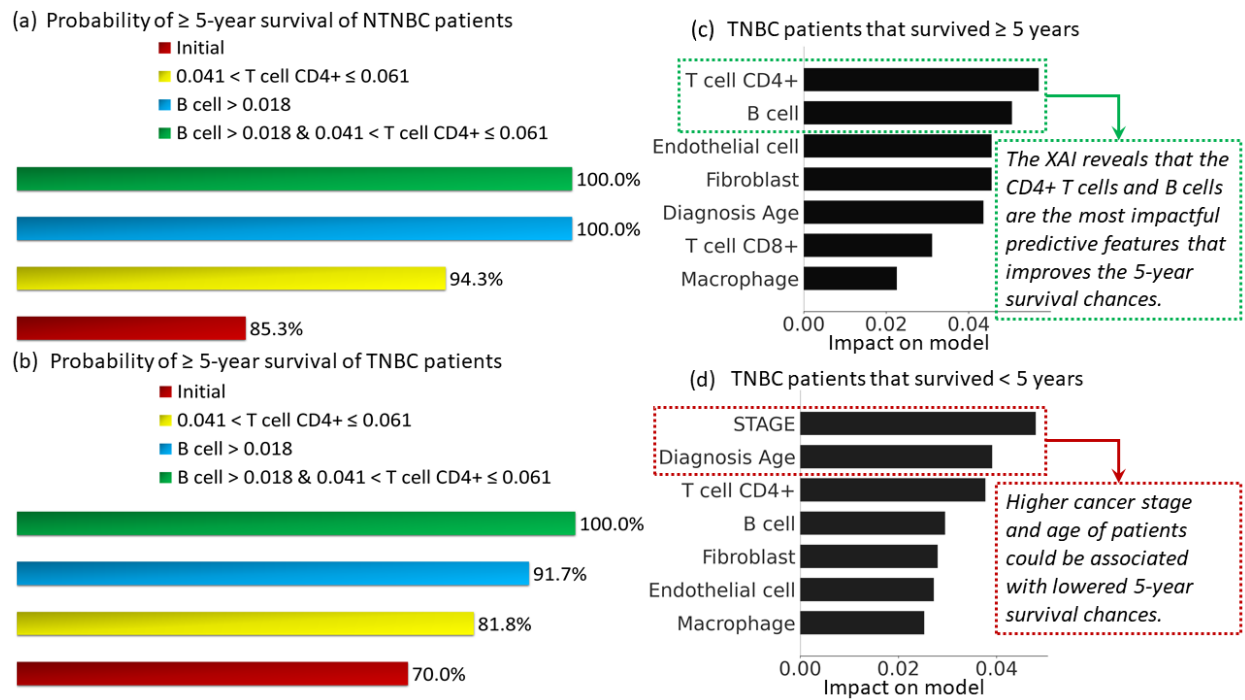

**Figure 5:** Conditional probabilities of ≥5-year survival of NTNBC **(a)** vs. TNBC **(b)** patients under various TME conditions. The initial data come from patients who are still alive after ≥5 years, are dead after surviving ≥5 years, or are dead after surviving <5 years. The most influential factors that are predictive of final outcome in TNBC patients who survived ≥5 years **(c)** and those who survived <5 years **(d)** are shown.

# 4 Discussion and Conclusion

Application of AI models for diagnostic and prognostic assessments has been widely accepted in the context of some cancers [40, 41]. The ability of AI models to discover embedded nonlinear patterns within complex multivariate datasets could potentially lead to a better understanding of the complex mechanisms that underlie carcinogenesis and cancer progression [42]. However, recent research indicates that there is a trend toward blind acceptance of black-box models that lack transparency and accountability, which could have severe consequences [43]. It is imperative to apply models that are inherently interpretable together with XAI methods to produce accurate predictions, to better understand the underlying reasoning of the AI approach, and to discover new interpretable knowledge from large datasets that would otherwise be impossible to detect with traditional statistical techniques [44]. To overcome such problems, which stem from the model's lack of transparency, we used XAI models comprising tree-based ensembles—which are more interpretable than the black box–type deep learning models (e.g., artificial neural networks)—along with game theory–based explanation models to determine prognoses in patients with breast cancer and to reveal valuable information regarding the ideal TME conditions for improved prognosis and treatments.

In the past, AI analysis has focused on early diagnosis of primary cancers, survival prognosis, or risk of relapse [45, 46]. Janizek et al. introduced "TreeCombo," a gradient of the boosted tree–based approach, in combination with Shapley analysis to predict synergy of novel drug combinations [47]. In various cancers, the density of tumor-infiltration lymphocytes (e.g., B cells, T cells) has correlated positively with survival prognosis [48, 49]. Using AI-based analysis (through a Random-Forest tree-based classifier) and CIBERSORT, He et al. [50] reported that High Immunity (a subset of TNBC according to the authors) was associated with larger numbers of CD8+ T cells, CD4+ T cells, NK cells, and M1 macrophages in the TME, and hence was considered to have more favorable clinical outcome than other subtypes of TNBC had. Similarly, with use of AI-based image analysis, He et al. [51] reported that tumor-infiltration lymphocytes cells were significantly reduced in the TME in metastatic TNBC, compared with the number of these cells in primary TNBC, and higher numbers of tumor-infiltration lymphocytes were found to be associated with better prognosis. Our XAI models indicated that maintaining the CD4+ T-cell and B-cell fractions in the TME within specific ranges would be conducive for optimal collaboration between T and B cells to carry out eradication of tumor cells, which may be related to CD4+ T cells causing B cells to proliferate and its progeny to differentiate into antibody-secreting cells. Subsequently, the B cells mark the tumor cells for destruction, which is carried out by cytotoxic cells such as CD8+ T cells and NK cells. Furthermore, this response is likely amplified by T-cell receptors arming the cytotoxic T cells. To the best of our knowledge, the relative importance and novel interactions between the tumor-infiltrating lymphocytes and tumor-associated cells in the TME on the overall and ≥5-year survival prognosis of breast cancer patients have not been previously reported, although such immune signatures could have potential clinical implications, especially for TNBC treatment.

Previous studies also reported that CD4+ and CD8+ T cells have opposing roles in breast cancer progression and outcomes: the CD8+ T cells were considered to be the crucial effector cells mediating effective antitumor immunity resulting in better clinical outcomes, whereas the intratumoral CD4+ T cells have negative prognostic effects on breast cancer patient outcomes [22]. Hollern et al. reported that CD4+ T-helper follicular cells, B cells, and the antibodies generated by those B cells play important roles in antitumor response to dual immune checkpoint inhibitors in mouse models [18, 52]. The authors observed prominent effects of CD4+ T-cell depletion and B-cell inhibition and/or depletion in mice, where therapeutic benefits were significantly reduced as assessed by survival and short-term response. Furthermore, previous use of rituximab to deplete B cells demonstrated no real clinical benefits for patients with solid tumors [53, 54]. The novel insights derived from our XAI model reinforce the finding that CD4+ T cells, along with B cells, are most critical in patients with breast cancer and reveal the critical inflection points of CD4+ T cells and B cells for designing innovative strategies to reprogram the TME, which results in improved prognoses in both TNBC and NTNBC patients. The main systemic therapy for metastatic TNBC is chemotherapy, despite the poor prognosis and poor patient survival associated with its use. In such cases, CD4+ T-cell and B-cell targeted immunotherapies could serve as a better alternative. Our findings underscore the need to rethink the current cancer immunotherapies focused on harnessing the antitumor CD8+ cytotoxic T-cell response. Increased awareness through use of XAI to understand the dynamics of the TME could lead to more rational and evidence-based therapies leading to improve outcomes in both TNBC and NTNBC patients.

## Authors Contributions

D.C., C.R-A., H.B., and G.L-B. conceived the study; H.B. and G.L-B. supervised the progress; D.C., C.R-A., H.B., and G.L-B. edited the manuscript; D.C., C.R-A., P.A., M.K., C. I., H.B., and G.L-B. designed and performed the experiments; D.C. and C.I., analyzed the data, and D.C., C.R-A., P.A., M.K., C. I., H.B., and G.L-B. wrote the manuscript.

## Acknowledgement

We thank Tamara Locke, Scientific Editor, Research Medical Library at The University of Texas MD Anderson Cancer Center for critical reading of the manuscript.

## Financial support

This work was supported in part by grants from the National Institutes of Health/National Cancer Institute (5U01CA213759-02, P30CA016672), and the American Cancer Society, National Science Foundation (CHE-1411859), and an endowment grant from the John P. Gaines Foundation. Dr. Cristian Rodriguez-Aguayo and Dr. Paola Amero were supported by the Brain SPORE Career Enhancement Program and NCI grant P50CA127001, as well as by the NIH through the Ovarian SPORE Career Enhancement Program and NCI grant P50CA217685.

## Conflicts of Interest

The authors declare no conflict of interest.